\documentclass[epj]{webofc}
\usepackage[utf8]{inputenc}
\usepackage[varg]{txfonts}   % Web of Conferences font
\usepackage{booktabs}
\usepackage{xcolor}
\definecolor{darkred}{rgb}{0.4,0.0,0.0}
\definecolor{darkgreen}{rgb}{0.0,0.4,0.0}
\definecolor{darkblue}{rgb}{0.0,0.0,0.4}
\usepackage[bookmarks,linktocpage,colorlinks,
    linkcolor = darkred,
    urlcolor  = darkblue,
    citecolor = darkgreen]{hyperref}
\usepackage{subfigure}
\wocname{EPJ Web of Conferences}
\woctitle{Lattice2017}
\begin{document}
%%%%%%%%%%%%%%%%%%%%%%%%%%%%%%%%%%%%%%%%%%%%%%%%%%%%%%%%%%%%%%%%%%%%%%%%%%%%%
%
\selectlanguage{english}
%----------------------------------------------------------------------------
\title{%
Higgs-Yukawa model on the lattice\thanks{Preprint number DESY 17-164}
}
%----------------------------------------------------------------------------
\author{%
\firstname{David Y.-J.} \lastname{Chu}\inst{1} \and
\firstname{Karl} \lastname{Jansen}\inst{2} \and
\firstname{Bastian} \lastname{Knippschild}\inst{3} \and
\firstname{C.-J. David}  \lastname{Lin}\inst{4}\fnsep\thanks{Speaker, \email{dlin@mail.nctu.edu.tw}}
% etc.
}
%----------------------------------------------------------------------------
\institute{%
Department of Electrophysics, National Chiao-Tung University, Hsinchu
30010, Taiwan
\and
NIC, DESY Zeuthen, D-15738, Germany
\and
HISKP and Bethe Centre for Theoretical Physics, University of Bonn,
D-53115, Germany
\and
Institute of Physics, National Chiao-Tung University, Hsinchu 30010, Taiwan
}
%----------------------------------------------------------------------------
\abstract{%
  We present results from two projects on lattice calculations for the
  Higgs-Yukawa model.  First we report progress on the search of
  first-order thermal phase transitions in the presence of a
  dimension-six operator, with the choices of bare couplings that lead
  to viable phenomenological predictions.  In this project the
  simulations are performed using overlap fermions to implement the
  required chiral symmetry.  Secondly, our study for applying finite-size scaling techniques near the Gaussian fixed point of the Higgs-Yukawa model is presented.   We discuss the analytical formulae for the Higgs Yukawa model and show results for a first numerical study in the pure $O(4)$ scalar sector of the theory.
}
%----------------------------------------------------------------------------
\maketitle
%----------------------------------------------------------------------------
\section{Introduction}\label{intro}

Investigation of the Higgs-Yukawa model using Lattice Field Theory can
result in input for physics at the Large Hadron Collider
(LHC).  In particular, it enables us to extend the study of the model
to the non-perturbative regime where a rich phase structure is being
found~\cite{Hasenfratz:1992xs, Gerhold:2007yb, Gerhold:2007gx, Bulava:2012rb}.  Such a phase structure can be employed to address the
hierarchy problem and the issue of triviality.  It can also provide
insight into the nature of electroweak thermal phase transition which
plays an important role in phenomenology of baryogenesis.

This article presents two projects on the Higgs-Yukawa model.  First
we discuss the extension of the theory with a dimension-six
operator that may lead to a strong first-order thermal phase
transition~\cite{Grojean:2004xa, Huang:2015izx, Damgaard:2015con,
  Cao:2017oez, deVries:2017ncy}.
Secondly, we show results from our study of finite-size scaling for
the Higgs-Yukawa model near the Gaussian fixed point.  Unlike the
usual practice in finite-size scaling where the scaling functions is
unknown, in this project we are able to derive these functions.  To
our knowledge this is the first time such finite-size scaling
functions are determined, and confronted with data from lattice
simulations\footnote{In Ref.~\cite{Gockeler:1992zj}, the authors
  derived similar scaling functions for the pure scalar $O(4)$ model in
the large-volume limit.}.  Such a study is
of its own interests and
will allow us to develop useful tools in looking for possible
non-trivial fixed points at strong coupling. 

In this work, we investigate the Higgs-Yukawa model that is described by the
continuum action
\begin{eqnarray}
\label{eq:cont_action}
 && S^{{\mathrm{cont}}}[\varphi,\bar{\psi},\psi] = \int \mathrm{d}^{4}x \left\{ \frac{1}{2} \left( \partial_{\mu}\varphi\right)^{\dagger}  
 \left( \partial_{\mu}\varphi  \right)
 + \frac{1}{2} m_{0}^2  \varphi^{\dagger} \varphi 
 +\lambda \left( \varphi^{\dagger} \varphi \right)^2 \right\} \nonumber \\
 && \hspace{3.0cm}+\int \mathrm{d}^{4}x \left\{ \bar{\Psi} \partial \hspace*{-1.8 mm} \slash \Psi 
 + y \left( \bar{\Psi}_{L} \varphi b_{R} + \bar{\Psi}_{L} \tilde{\varphi}
    t_{R} + h.c. \right) \right\}, \nonumber\\
 &&\hspace{0.5cm}{\mathrm{where}} \quad \varphi = \left ( \begin{array}{c}  \phi_2+i\phi_1 \\ 
                     \phi_0 - i\phi_3 \end{array} \right ), \quad
  \tilde{\varphi}=i\tau_2\varphi, \quad
\Psi = \left ( \begin{array}{c} t \\ b \end{array} \right ) , \quad
  \Psi_{L,R} = \frac{1\mp \gamma_{5}}{2} \Psi, 
\end{eqnarray}
with $\phi_{i}$ being real scalar fields, $t$ and $b$ being the ``top''
and the ``bottom'' quark fields, and $\tau_2$ being the second Pauli
matrix.  Amongst the scalar fields, the component $\phi_{0}$ will
develop a non-vanishing vacuum expectation value (vev) in the
phase of spontaneously-broken $O(4)$ symmetry.  The above action contains three bare couplings, $m_{0}$,
$\lambda$ and $y$.  Notice that we employ degenerate Yukawa couplings
in this work.
To discretise this action, we resort to overlap fermions that
allow us to properly define the lattice version of the left- and
right-handed fermions in the Yukawa terms.  Furthermore, we follow the
convention in representing the bosonic component of the lattice action as
\begin{equation}
S_B[\Phi] = -\kappa \sum\limits_{x,\mu} \Phi_x^{\dagger}
\left[\Phi_{x+\mu} + \Phi_{x-\mu}\right] + \sum\limits_{x}
\Phi_x^{\dagger} \Phi_x + \hat{\lambda}\sum\limits_{x} \left[
  \Phi_x^{\dagger} \Phi_x - 1 \right]^2 ,
\end{equation}
where $\kappa$ is the hopping parameter, $x$ labels lattice sites, and $\mu$ specifies the
space-time directions.  The relationship between the lattice and the
continuum bosonic fields and relevant couplings is
\begin{equation}
a \varphi = \sqrt{2 \kappa}  \Phi_{x} \equiv \sqrt{2 \kappa}
\left( \begin{array}{c} \Phi_{x}^2 + i\Phi_{x}^1 \\ \Phi_{x}^0 - i
         \Phi_{x}^3 \end{array} \right) ,\quad \lambda =
     \frac{\hat{\lambda}}{{4 \kappa^2}},\quad m_0^2 = \frac{1 - 2
       \hat{\lambda} -8 \kappa}{\kappa} , 
\end{equation}
where $a$ denotes the lattice spacing.

All the numerical works reported in this article have been performed
with the choice of the bare Yukawa coupling,
\begin{equation}
\label{eq:choice_of_bare_y}
 y = 175/246 ,
\end{equation}
as motivated by the physical values of the Higgs-field vev and the top-quark mass.

%----------------------------------------------------------------------------
\section{Thermal phase transition in the Higgs-Yukawa model with a
  dimension-six operator}\label{sec:dim6_op}

In this section, we present our study of the Higgs-Yukawa model with the 
inclusion of a term,
\begin{equation}
\label{eq:dim6_op}
 {\mathcal{O}}_{6} = \frac{\lambda_{6}}{\Lambda^{2}} \int
 {\mathrm{d}}^{4}x \left ( \varphi^{\dagger} \varphi \right )^{3} ,
\end{equation}
in the action of Eq.~(\ref{eq:cont_action}).  This term can serve
as a prototype of new physics.  Here
$\Lambda$ is the cut-off scale that can be realised as $1/a$ on
the lattice.  It is natural to include
this operator as one of the higher-dimension terms when interpreting
the Higgs-Yukawa model in the language of effective field theory~\cite{Bilenky:1994kt}.
In the above expression, $\lambda_{6}$ is
dimensionless.  The addition of the dimension-six operator, $\left (
  \varphi^{\dagger} \varphi \right )^{3}$, can enrich both thermal and
non-thermal phase structures of the theory. 

Our main interest in the Higgs-Yukawa model with the operator in
Eq.~(\ref{eq:dim6_op}) is the search for a viable scenario for a
strong first-order thermal phase transition in the theory, while
maintaining a second-order transition at zero temperature.  In performing such
search, we scan the bare-coupling space to identify choices of
parameters such that
\begin{enumerate}
 \item The cut-off is high enough compared to the renormalised Higgs-field vev, 
    denoted as $\langle \varphi \rangle \equiv \langle \phi_{0} \rangle$.  This means the condition,
\begin{equation}
\label{eq:high_cutoff}
  a \langle \varphi \rangle \ll 1 ,
\end{equation}
 has to be satisfied in our simulations.
 \item The ratio between $\langle \varphi \rangle$ and the Higgs-boson mass is compatible
   with experimental results,
   {\it i.e.},
\begin{equation}
\label{eq:higgs_vev_mass_ratio}
 \frac{\langle \varphi \rangle}{m_{H}} \sim 2 .
\end{equation}
 \item The thermal phase transition is first-order.
\end{enumerate}
To ensure that Eq.~(\ref{eq:high_cutoff}) is realised in our
simulations, we have to examine the non-thermal phase structure of the
model.  In the phase where the $O(4)$ symmetry is spontaneously
broken, this condition can be satisfied near any second-oder
non-thermal phase transitions.  In our previous
work~\cite{Chu:2015nha}, a thorough investigation in this regard  was
conducted for two choices of $\lambda_{6}$
($\lambda_{6} = 0.001$ and $\lambda_{6} = 0.1$), leading to useful
information for the current study.  In order to check the constraint of
Eq.~(\ref{eq:higgs_vev_mass_ratio}), we determine the Higgs-boson mass
from the momentum-space Higgs propagator.  

An important tool in our study of the phase structures is the
constraint effective potential (CEP)~\cite{Fukuda:1974ey,
  ORaifeartaigh:1986axd}\footnote{An alternative, somewhat similar,
  tool for such studies is the extended mean-field
  theory~\cite{Akerlund:2015fya}.}.  The CEP, $U(\hat{v})$, is a
function of the Higgs-field zero mode,
\begin{equation}
 \hat{v} = \frac{1}{V} \left | \sum_{x} \Phi^{0}_{x} \right |,
\end{equation}
where $V$ is the four-volume and the sum is over all lattice points.
This effective potential can
be calculated analytically using perturbation theory.  It can also be
extracted numerically  through a histogramming procedure of $\hat{v}$ in
Monte-Carlo simulations.

\begin{figure}[thb] 
  \centering
  \includegraphics[width=7.5cm,clip]{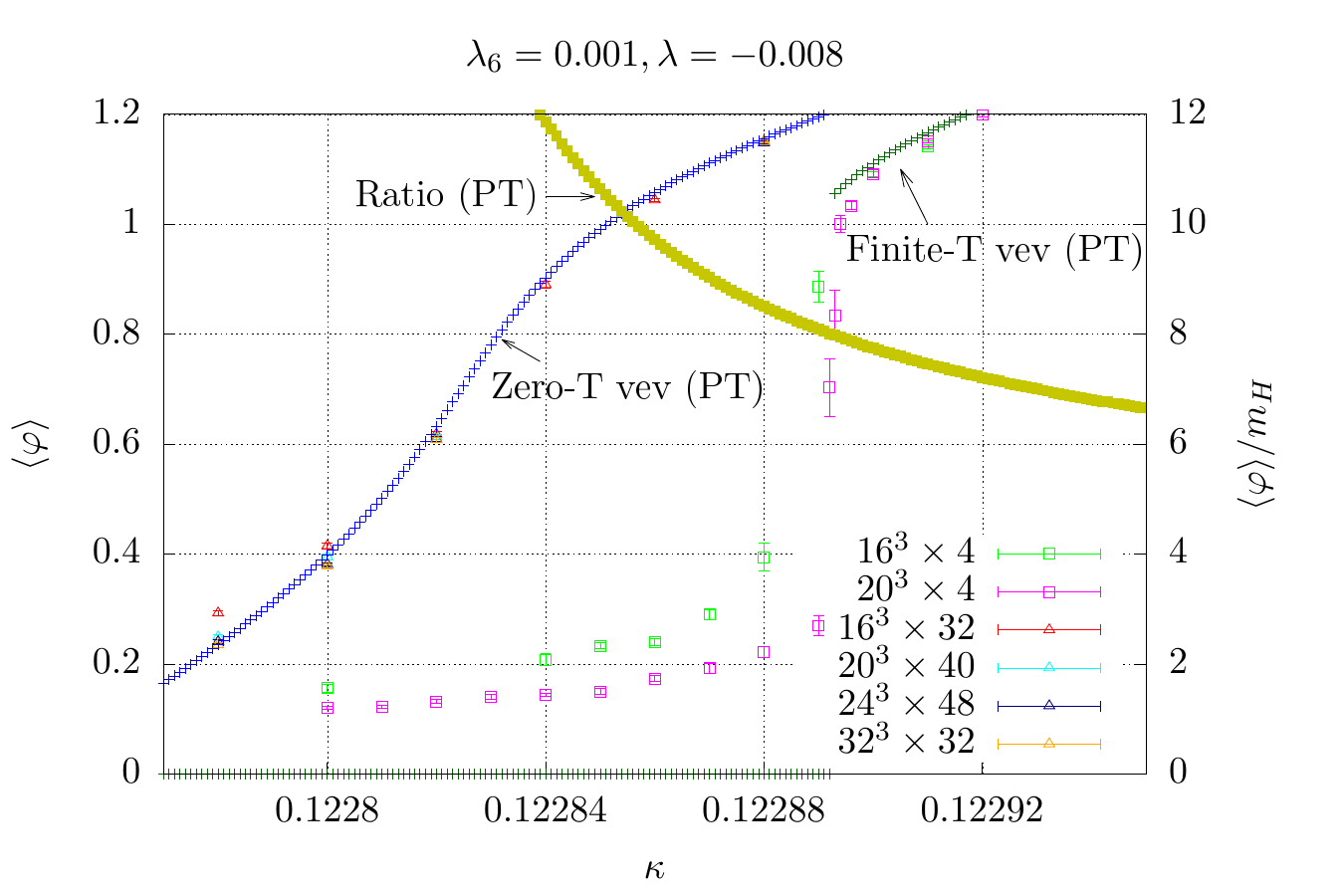}
  \includegraphics[width=6.5cm,clip]{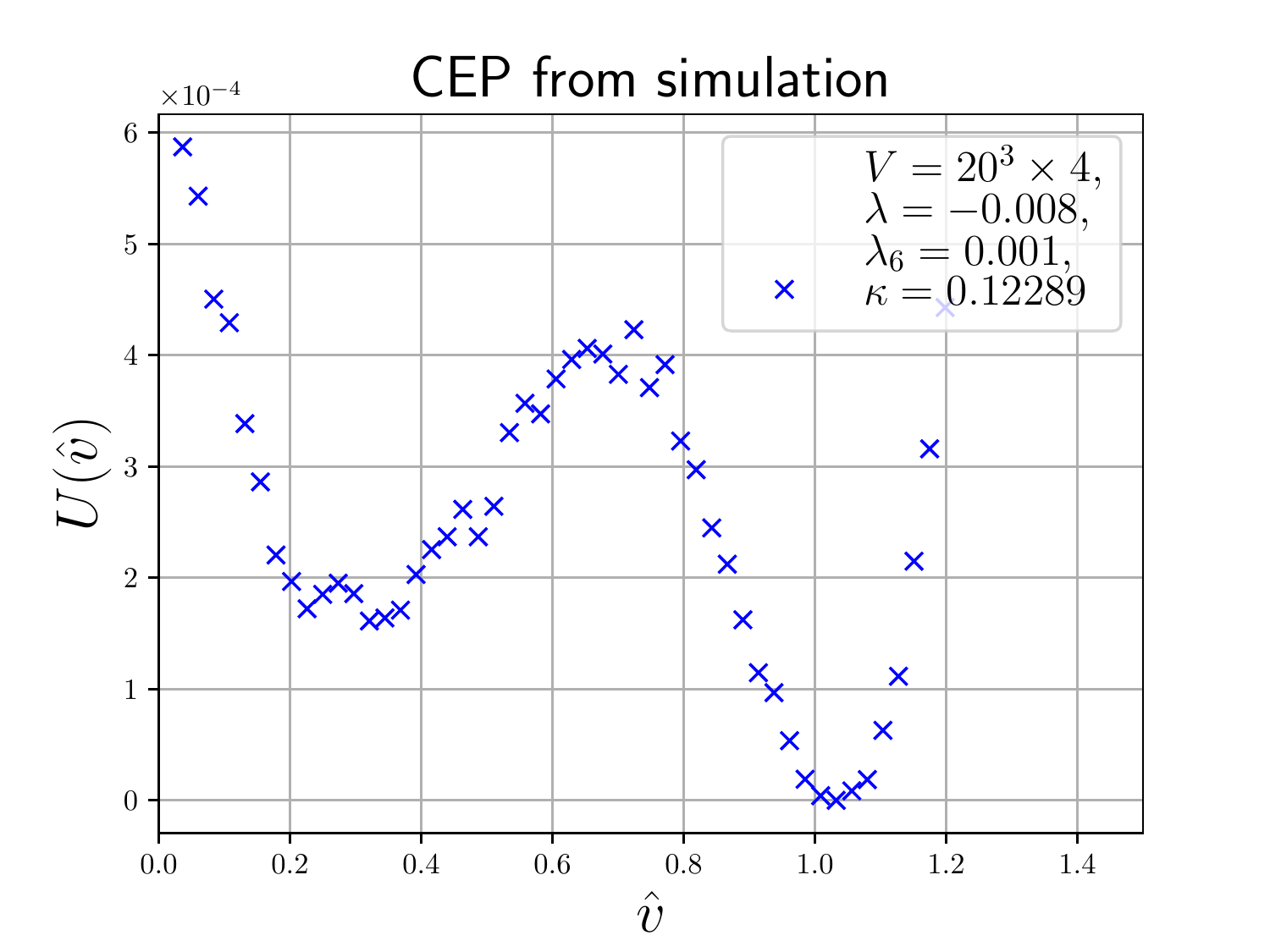}
  \caption{Results for $\lambda_{6} =
    0.001$ and $\lambda = -0.008$. In the left-hand panel, the errors
    are statistical from lattice computations, and symbols without
    errors represent results obtained in perturbative calculations
    using the constraint effective potential.  ``Ratio'' means
    $\langle \varphi \rangle/m_{H}$, and ``PT'' stands for
    ``perturbation theory''. In this figure, the Higgs vev,
    $\langle \varphi \rangle$,
    is plotted in lattice units.  The right-hand side is the
    constraint effective potential, obtained in the Monte-Carlo
    simulation at non-zero temperature, and at the four-volume $(L/a)^{3}\times (T/a) = 20^{3} \times 4$ and $\kappa =
    0.12289$.}
  \label{fig:CEPlambda6euqls0point001}
\end{figure}

Figure~\ref{fig:CEPlambda6euqls0point001} exhibits our results at
$\lambda_{6} = 0.001$ and $\lambda = -0.008$.  
According to the
study in Ref.~\cite{Chu:2015nha}, perturbative calculations of the
CEP are reliable at these values of the couplings.  The plot in the
left-hand panel demonstrates that this choice of the self
couplings can lead to a second-order non-thermal and a first-order
thermal phase transitions.  The first-order transition is further
evidenced by our numerical study of the CEP, as presented in the
right-hand panel.  On the other hand, results from perturbation theory
show that the ratio $\langle \varphi \rangle /m_{H}$ does not satisfy
the condition in Eq.~(\ref{eq:higgs_vev_mass_ratio}).

\begin{figure}[thb] 
  \centering
  \includegraphics[width=4.65cm,clip]{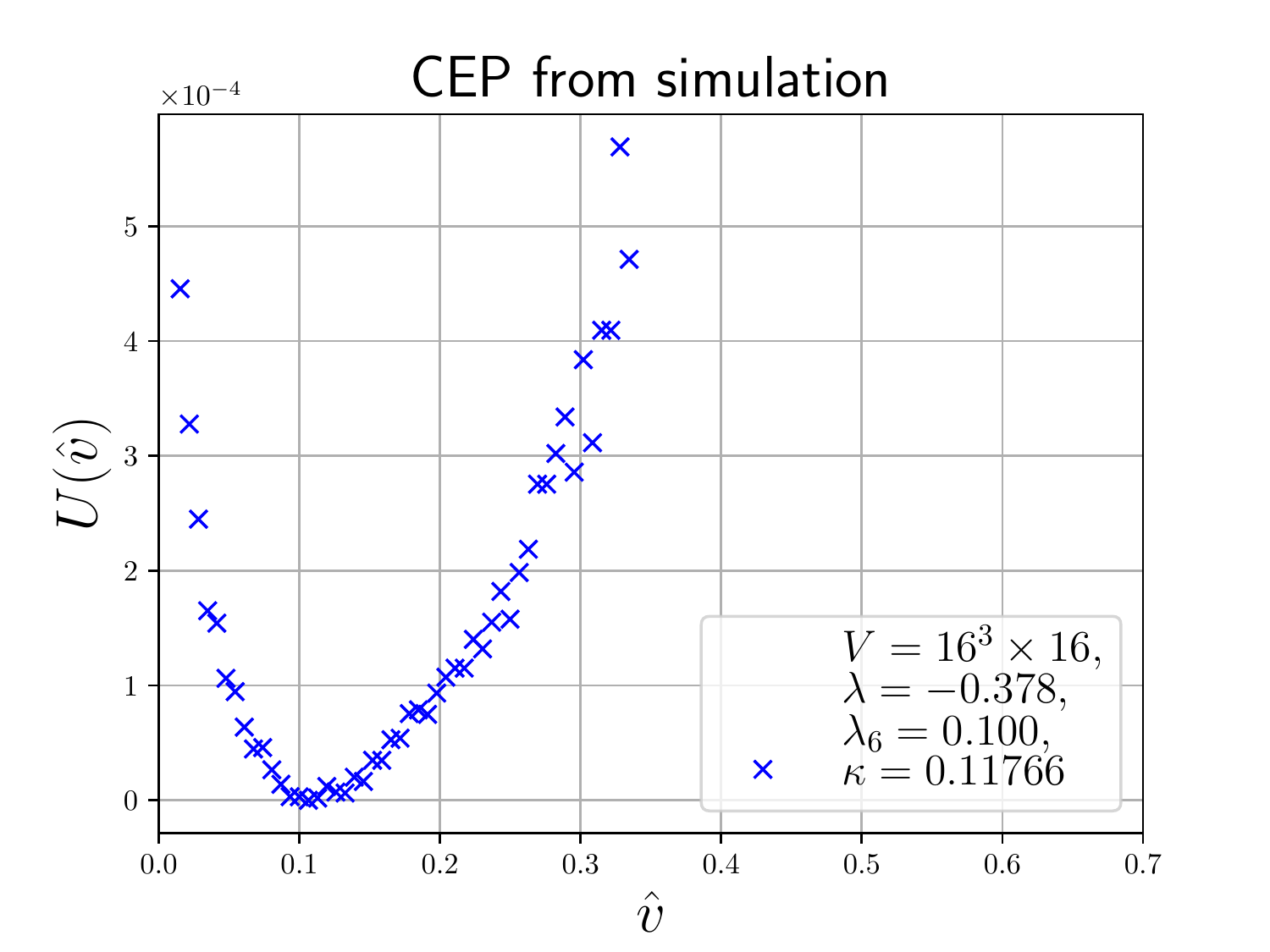}
  \includegraphics[width=4.65cm,clip]{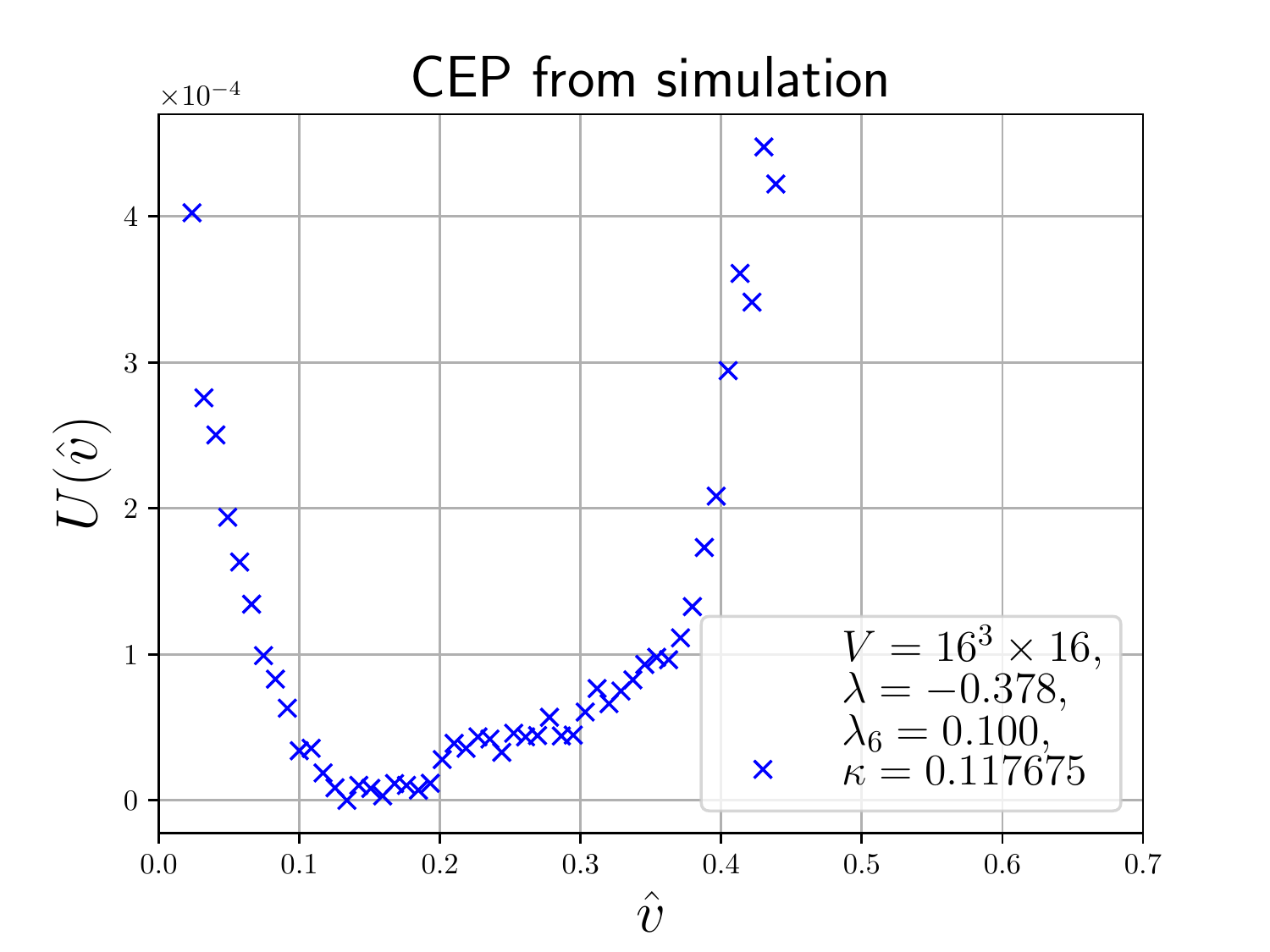}
  \includegraphics[width=4.65cm,clip]{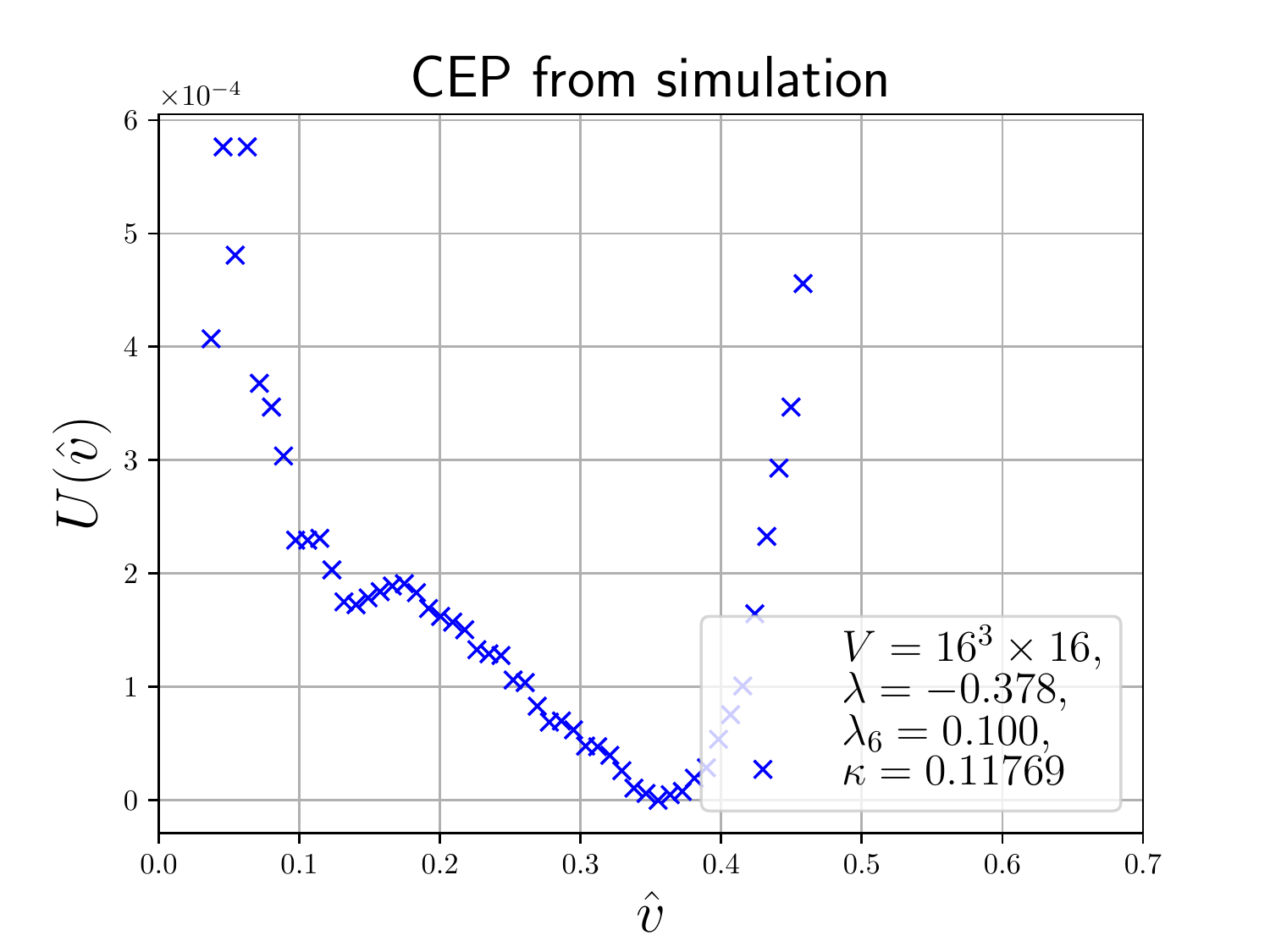}\\
  \includegraphics[width=4.65cm,clip]{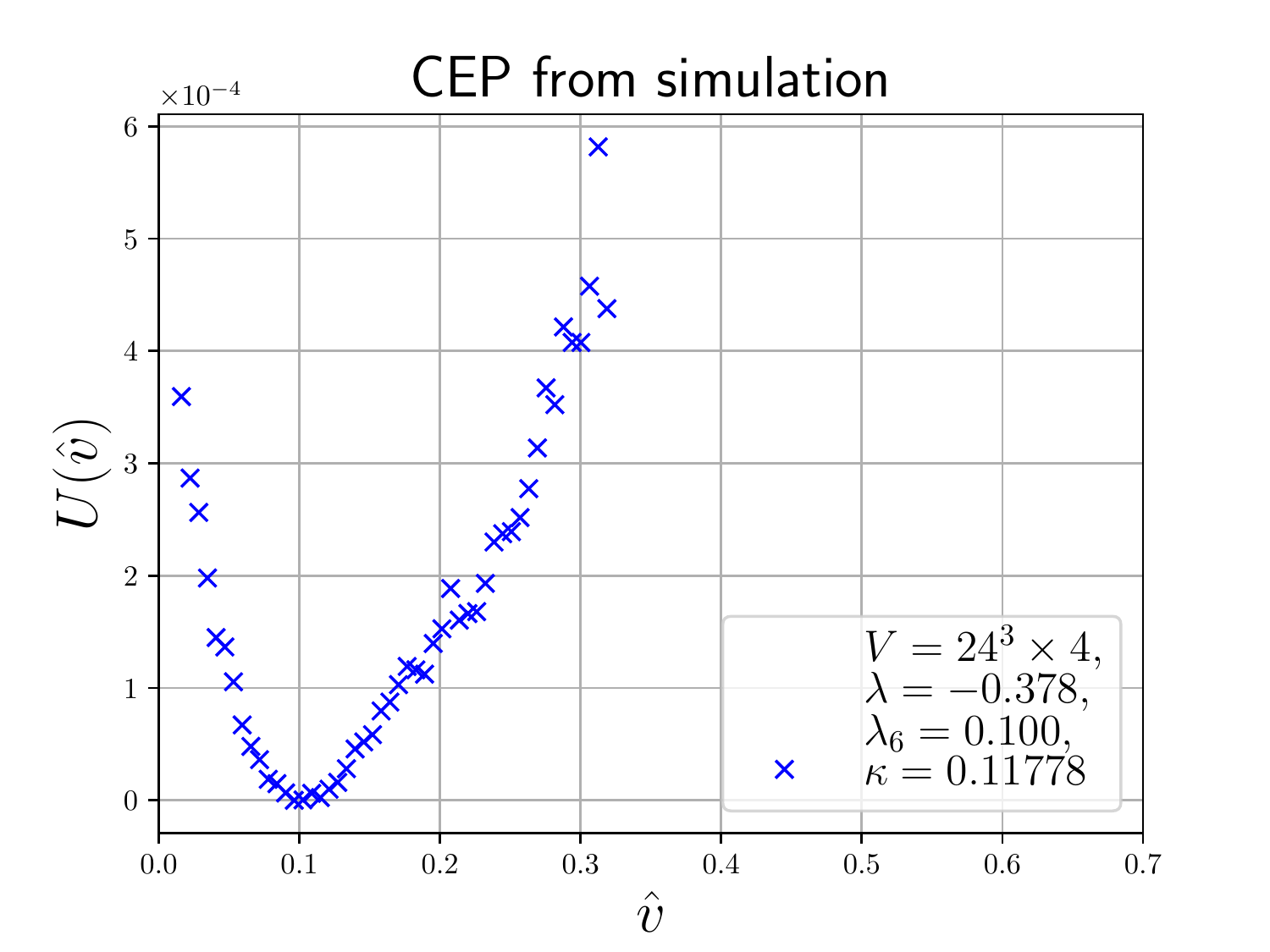}
  \includegraphics[width=4.65cm,clip]{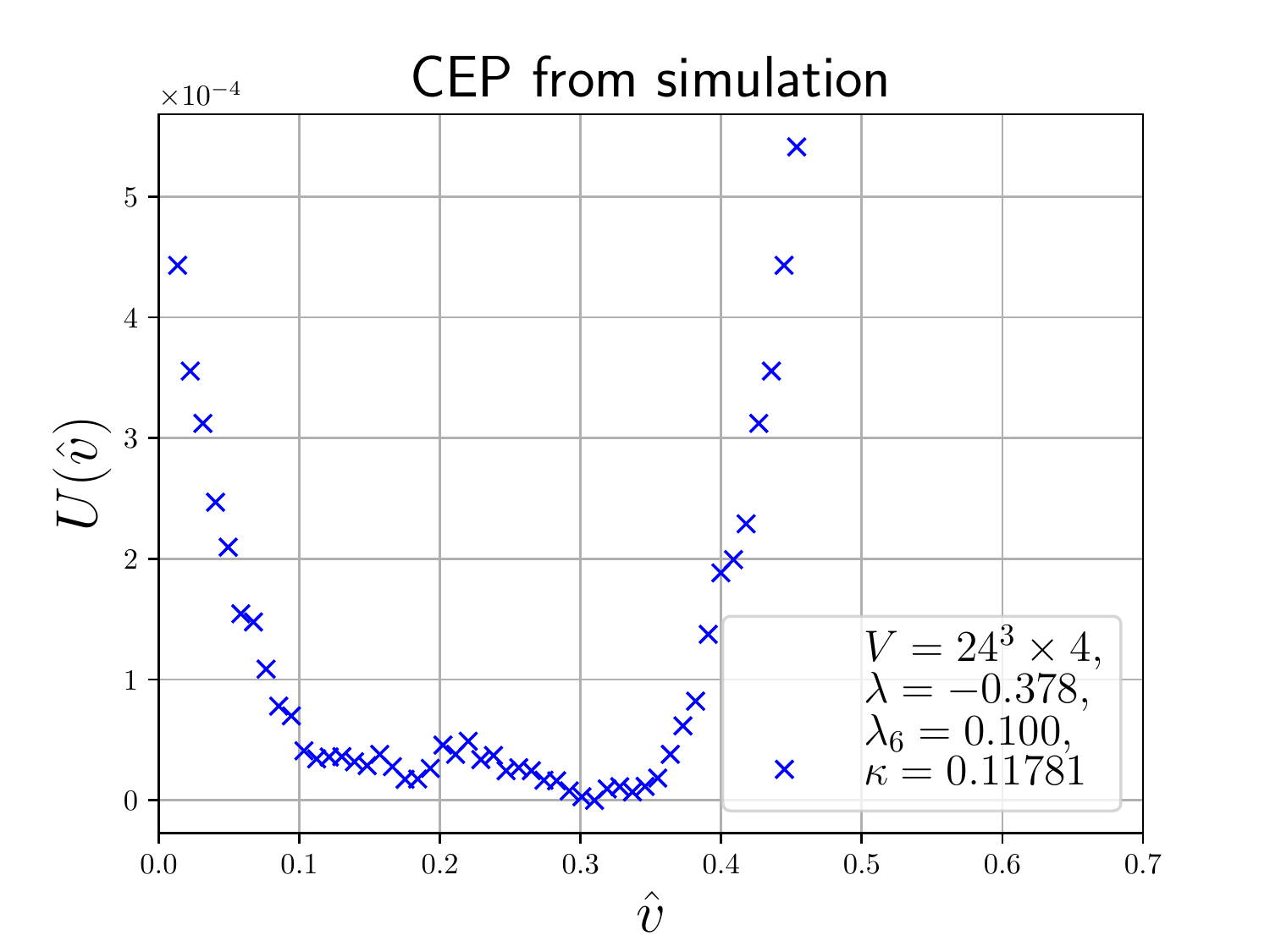}
  \includegraphics[width=4.65cm,clip]{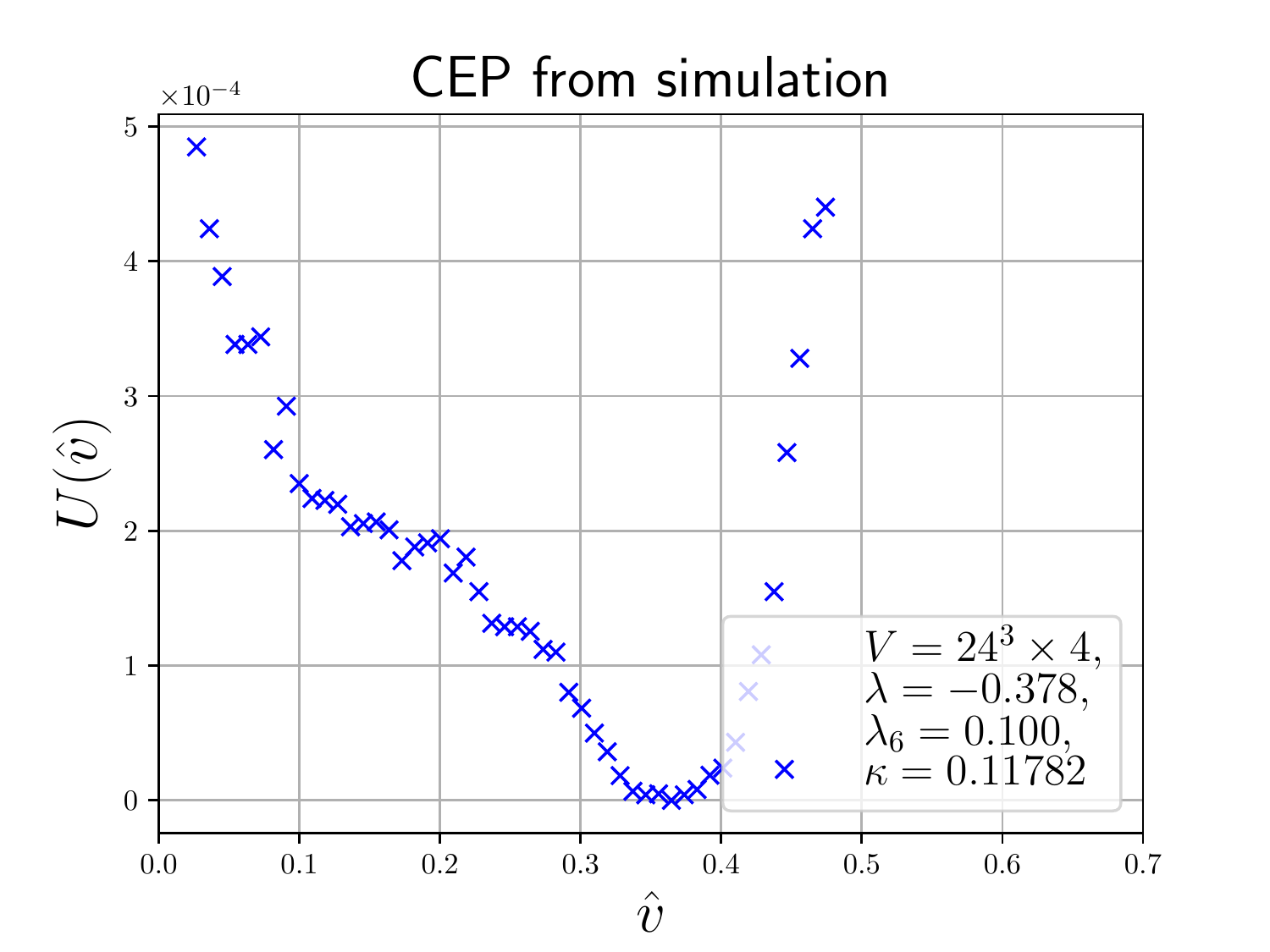}
  \caption{Results for $\lambda_{6} =
    0.1$ and $\lambda = -0.378$.  The plots on the first row are the
    CEP at three representative values of $\kappa$ near the
    non-thermal phase transition.  The second row displays their
    counterparts at the finite-temperature transition.}
  \label{fig:CEPlambda6euqls0point1}
\end{figure}

To perform further search, we choose $\lambda_{6} = 0.1$ and
$\lambda=-0.378$.   Our work in Ref.~\cite{Chu:2015nha} shows that
perturbation theory for the CEP is no longer reliable at this value of
$\lambda_{6}$.  Therefore we only resort to Monte-Carlo simulations on
the lattice.  Results for the CEP near the phase transitions are displayed in
Fig.~\ref{fig:CEPlambda6euqls0point1}.  
 Although it is not clear
whether these transitions are first- or second-order, it can be
concluded that the non-thermal and the thermal phase transitions
exhibit almost the same properties.  Hence we conclude that this
choice of the bare scalar self-couplings does not lead to a viable
scenario for a strong first-order thermal phase transition.

Presently we are continuing this search with other choices of
$\lambda_{6}$ and $\lambda$.
%%
%\begin{table}[thb]
%  \small
%  \centering
%  \caption{Please write your table caption here}
%  \label{tab-1}% Give a unique label
%  \begin{tabular}{lll}\toprule
%  first  & second & third  \\\midrule
%  number & number & number \\
%  number & number & number \\\bottomrule
%  \end{tabular}
%\end{table}
%%

%----------------------------------------------------------------------------
\section{Finite-size scaling for the Higgs-Yukawa model near the
  Gaussian fixed point}
\label{sec:FFS}
In the second project, we study the Higgs-Yukawa model as described by the
continuum action in Eq.~(\ref{eq:cont_action}), {\it i.e,}, we have
\begin{equation}
 \lambda_{6} = 0 ,
\end{equation}
and only include operators with dimension less than or equal to four in the
action.  The purpose of this investigation is to develop tools for
exploring the scaling behaviour of the model near the Gaussian fixed
point.  These tools can be used to confirm the triviality of the
Higgs-Yukawa theory, or to search for alternative scenarios where
strong-coupling fixed points exist.   Predictions from perturbation theory indicate the
possible appearance of non-trivial fixed points in the
Higgs-Yukawa model~\cite{Molgaard:2014mqa}.  This issue was also examined with the
approach of functional renormalisation group~\cite{Gies:2009hq}, where
no non-Gaussian fixed point was found.   Nevertheless, early lattice
computations showed evidence for the opposite conclusion~\cite{Hasenfratz:1992xs}.
We stress that an extensive, non-perturbative study of the
Higgs-Yukawa theory from first-principle calculations is
still wanting.   This is in contrast to the situation of the
pure-scalar models which are now widely believed to be 
trivial (see, {\it e.g.}, Refs.~\cite{Frohlich:1982tw, Luscher:1988uq,
  Hogervorst:2011zw,
Siefert:2014ela}).

In our previous attempt at addressing the issue of the triviality in the
Higgs-Yukawa model, as reported in Ref.~\cite{Bulava:2012rb}, we employed the technique of
finite-size scaling.  The main finding in Ref.~\cite{Bulava:2012rb} is
that one needs to understand the logarithmic corrections to the
mean-field scaling behaviour, in order to draw concrete conclusions.
In view of this, we developed a strategy, and worked out its analytic
aspects, as reported in Refs.~\cite{Chu:2015jba, Chu:2016svq}. 

The main result in Refs.~\cite{Chu:2015jba, Chu:2016svq} is that, using the
techniques established by Zinn-Justin and Brezin for scalar field
theories~\cite{Brezin:1985xx}, we can derive finite-size scaling
formulae for various quantities in
one-loop perturbation theory near the Gaussian fixed point.  It is natural to include the
leading-order logarithmic corrections to the mean-field scaling law through
this procedure.   In this strategy, we first match
correlators obtained with lattice regularisation to an on-shell renormalisation
scheme, with the matching scale chosen to be the pole mass, $m_{P}$, of the
scalar particle.  This pole mass can be extracted by studying the
scalar-field propagator on the lattice.  Its relationship with the
renormalised mass parameter [the renormalised counterpart of the
bare coupling $m_{0}$ in Eq.~(\ref{eq:cont_action})], $m$, in the theory is
\begin{eqnarray}
\label{eq:pole_to_renorm_mass}
 && m^2 (m_P) =  m_{P}^{2} \mbox{ }\mbox{ }{\mathrm{in}}\mbox{
               }{\mathrm{the}}\mbox{ }{\mathrm{symmetric}}\mbox{ }{\mathrm{phase}} , \nonumber \\
 && m^2 (m_P) = -\frac{1}{2} m_{P}^{2} \mbox{ }\mbox{ }{\mathrm{in}}\mbox{
               }{\mathrm{the}}\mbox{ }{\mathrm{broken}}\mbox{ }{\mathrm{phase}} ,
\end{eqnarray}
where the renormalisation scale is $m_{P}$.  Notice that $m_{P}$ is
the Higgs-boson pole mass in the broken phase.  
Under the assumption that we
work closely enough to the critical surface of the Gaussian fixed
point, the condition $m_{P} \ll 1/a$ is satisfied.  We can then carry out one-loop running of the renormalised correlators from
$m_{P}$ to another low-energy scale that is identified with the
inverse lattice size, $L^{-1}$, that is of the same order of $m_{P}$
but with the constraint $m_{P} L > 1$.  This leads to predictions of finite-size scaling
behaviour of these correlators.  In performing the above one-loop running, one has to
solve the relevant renormalisation group equations, introducing integration
constants.  These constants will then be treated as fit parameters
when confronting the scaling formulae with lattice numerical data.  

Up to the effect of wavefunction renormalisation, which results in
additional $L{-}$dependence that can also be accounted for using one-loop
perturbation theory, we have found that all the correlators containing
only the zero mode of the scalar field can be expressed in terms of a
class of functions
\begin{eqnarray}
% \label{first_integral}
 \bar{\varphi}_0 (z) &=& \frac{\pi}{8} \exp \left( \frac{z^2}{32} \right) \sqrt{|z|} 
 \left[ I_{-1/4}\left( \frac{z^2}{32} \right) - \mathrm{Sgn}(z) \, I_{1/4}\left( \frac{z^2}{32} \right)  \right],\nonumber \\
% \label{second_integral}
 \bar{\varphi}_1 (z) &=& \frac{\sqrt{\pi}}{8} \exp \left( \frac{z^2}{16} \right) 
 \left[ 1-\mathrm{Sgn}(z) \, \mathrm{Erf} \left( \frac{|z|}{4} \right) \right], \mbox{ } 
%  \nonumber \\
 \label{recursion_formula}
 \bar{\varphi}_{n+2} (z) = -2 \frac{\mathrm{d}}{\mathrm{d} z}
 \bar{\varphi}_n (z), 
\end{eqnarray}
where the scaling variable, $z$, is 
\begin{equation}
\label{scaling_variable}
 z=\sqrt{s}m^{2} \left( L^{-1} \right) L^{2} \lambda_{R} \left( L^{-1} \right)^{-1/2},
\end{equation}
with $\lambda_{R} (L^{-1})$ being the quartic coupling renormalised at
the scale $L^{-1}$, and $s$ being the anisotropy of the four-volume,
$L^{3}\times T = L^{3} \times sL$.  Notice that the $L{-}$dependence
in $\lambda_{R} (L^{-1})$ can be complicated, involving two
integration constants resulted from solving the renormalisation group equations for the
Yukawa and the quartic couplings~\cite{our_scaling_paper}.  On the
other hand, there is no unknown parameter in the renormalised mass,
$m(L^{-1})$, because it is obtained from the scalar pole mass computed
numerically on the lattice.  To our knowledge this is the first time the formulae in
Eq.~(\ref{recursion_formula}) are derived, although similar results
were obtained in the large-volume limit in Ref.~\cite{Gockeler:1992zj}.

In this work, we study the
Higgs-field vev, $\langle \varphi \rangle \equiv \langle \phi_{0} \rangle$, its susceptibility, $\chi$, and Binder's cumulant,
$Q$.  The finite-size scaling formulae for these quantities are found
to be
\begin{eqnarray}
 \label{eq:scaling_formulae}
  \langle \varphi \rangle
     &=& s^{-1/4} A^{(\varphi)} L^{-1} \left[ \lambda_{R} (L^{-1}) \right]^{-1/4}
     \frac{\bar{\varphi}_4(z)}{\bar{\varphi}_3(z)}, \nonumber \\ 
    \chi &=& s L^4 \left( \langle \varphi^2 \rangle -\langle \varphi \rangle^2 \right)
     = s^{1/2} A^{(\chi)} L^{2} \left[ \lambda_{R} (L^{-1}) \right]^{-1/2}
     \left[ \frac{\bar{\varphi}_5(z)}{\bar{\varphi}_3(z)} - 
     \left( \frac{\bar{\varphi}_4(z)}{\bar{\varphi}_3(z)} \right)^2
            \right], \nonumber \\
   Q &=& 1-\frac{\langle \varphi^4 \rangle}{3\langle \varphi^2 \rangle^2} 
 =
 1-\frac{\bar{\varphi}_7(z)\bar{\varphi}_3(z)}{3\bar{\varphi}_5(z)^2}
 , 
\end{eqnarray}
where $\langle \varphi^{2} \rangle  \equiv \langle \phi_{0}^{2}
\rangle$ in the definition of the susceptibility, $A^{(\varphi)}$ and
$A^{(\chi)}$ are unknown constants resulting from integrating the
renormalisation group equation for the the wavefunction.

As mentioned above, the formulae in Eq.~(\ref{eq:scaling_formulae})
can be complicated in the Higgs-Yukawa theory.  Therefore as the first
numerical test of our strategy, we resort to the pure-scalar $O(4)$
model, in which the one-loop $\lambda_{R} (L^{-1})$ takes the simple form,
\begin{equation}
\label{eq:lambda_R_O4}
 \lambda_{R} (L^{-1}) =
 \frac{\lambda_{m_{P}}}{1+\frac{6}{\pi^2}\log(m_{P}L)} ,
\end{equation}
with only one integration constant, $\lambda_{m_{P}} \equiv \lambda_{R} (m_{P})$, to be
fitted from numerical data.  In this numerical test, the bare quartic
coupling is chosen to be 0.15, to ensure that we are working in the
perturbative regime.  Throughout our analysis procedure, the
scalar-particle pole mass is determined by fitting the four-momentum
space propagator, and then extrapolated to the infinite-volume limit
employing a ChPT-inspired formula.  More details of this aspect of our
work will be reported in a near-future publication~\cite{our_scaling_paper}.

\begin{figure}[thb] 
  \centering
  \includegraphics[width=4.65cm,clip]{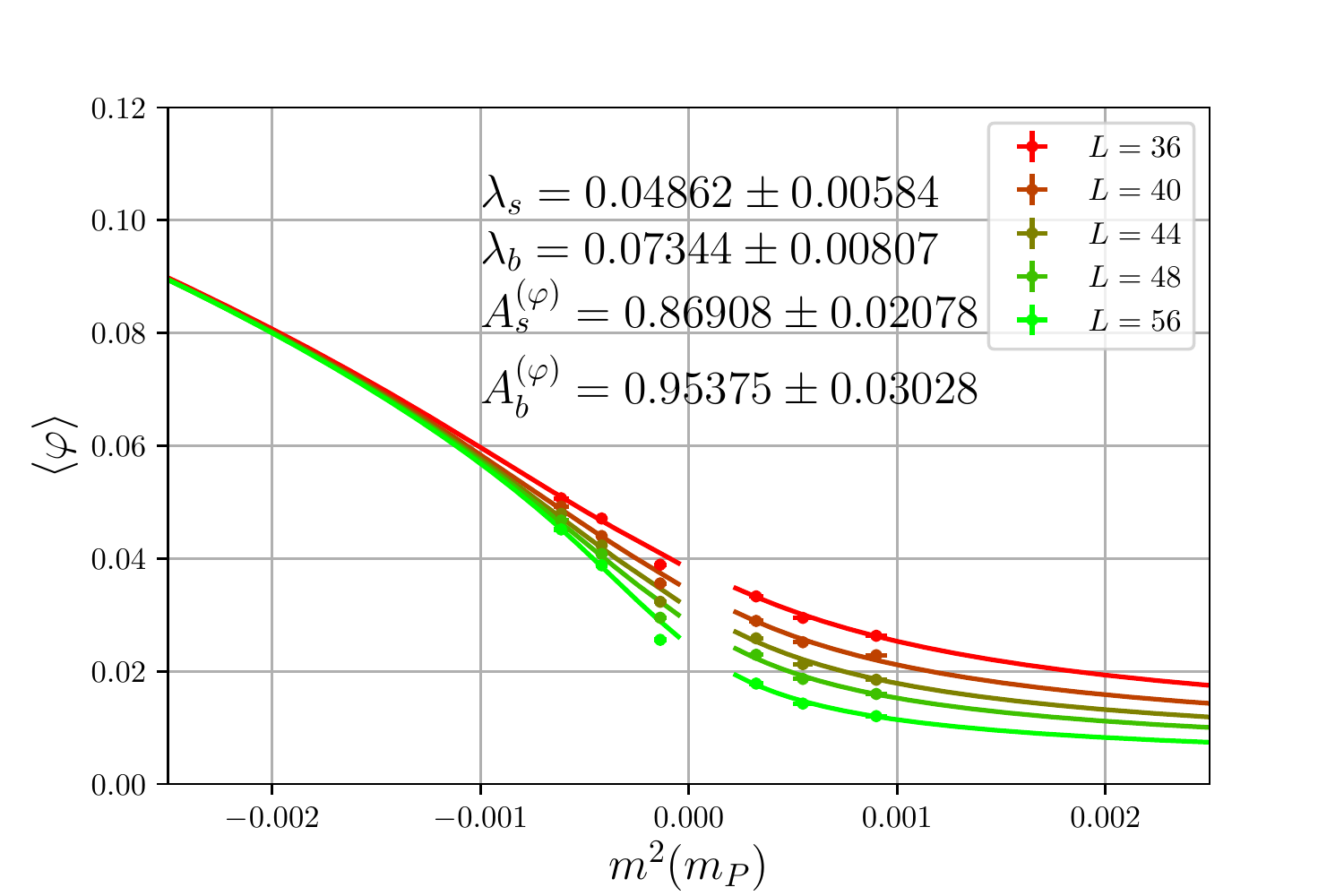}
  \includegraphics[width=4.65cm,clip]{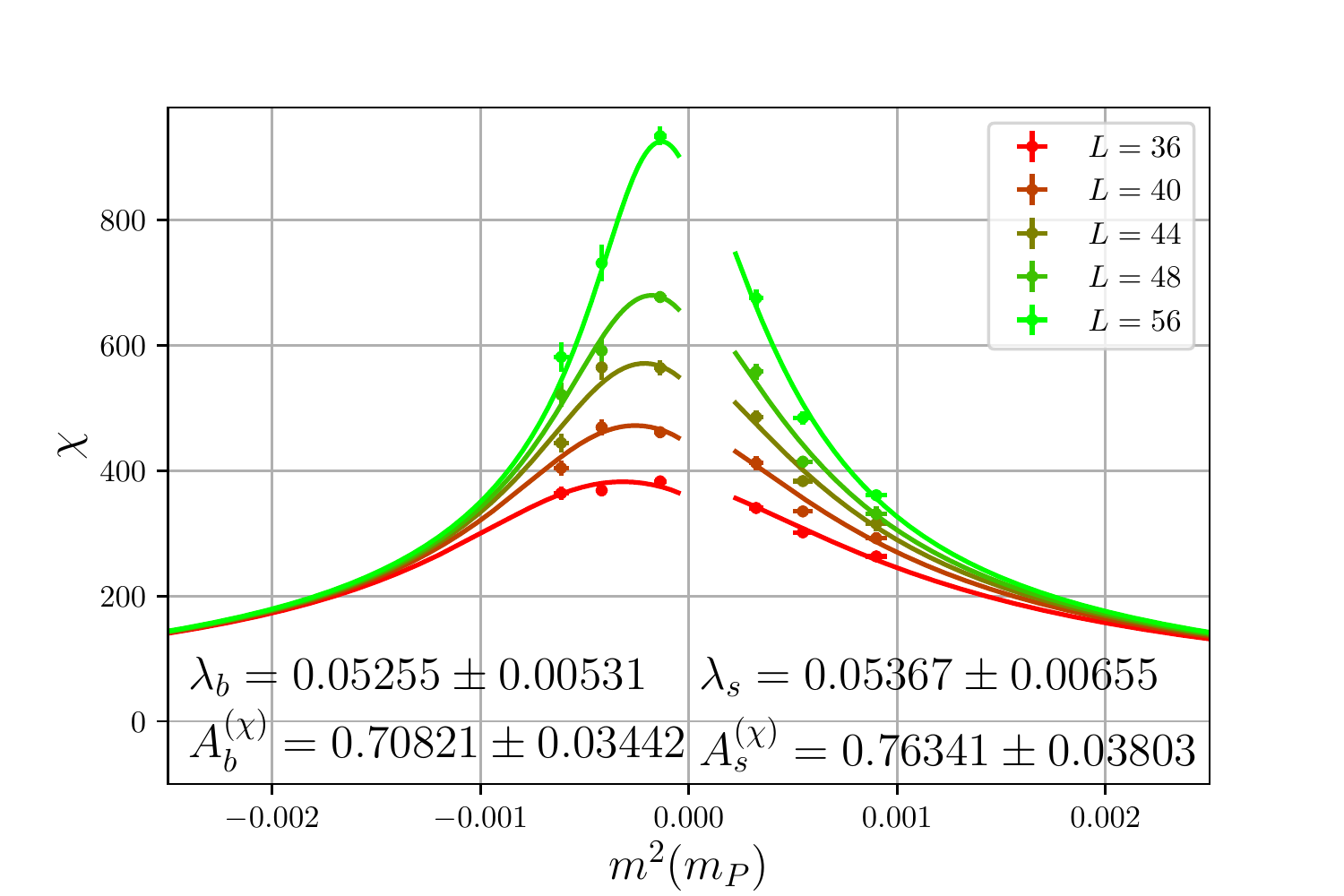}
  \includegraphics[width=4.65cm,clip]{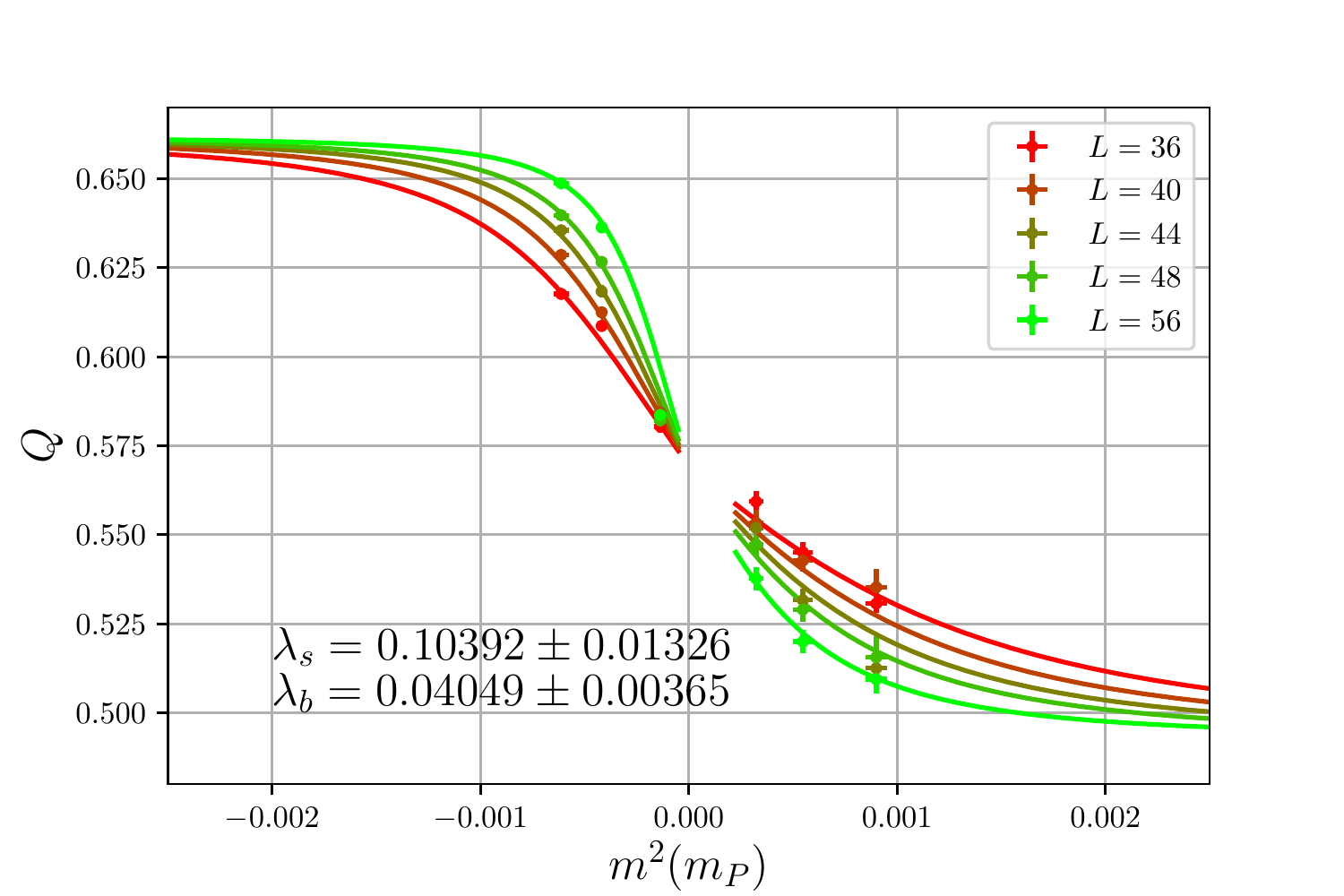}
  \caption{Results for the fit of the Higgs-field vev, its
    susceptibility and Binder's cumulant using the finite-size scaling
  formulae in the pure-scalar $O(4)$ model.  The parameters
  $\lambda_{s,b}$ are $\lambda_{m_{R}}$ in the symmetric and the
  broken phases, with similar symbols indicating the relevant
  $A^{(\varphi)}$ and $A^{(\chi)}$ in these two phases.   All dimensionfull
  quantities are expressed in lattice units.}
  \label{fig:Fit_scaling}
\end{figure}

\begin{figure}[thb] 
  \centering
  \includegraphics[width=4.65cm,clip]{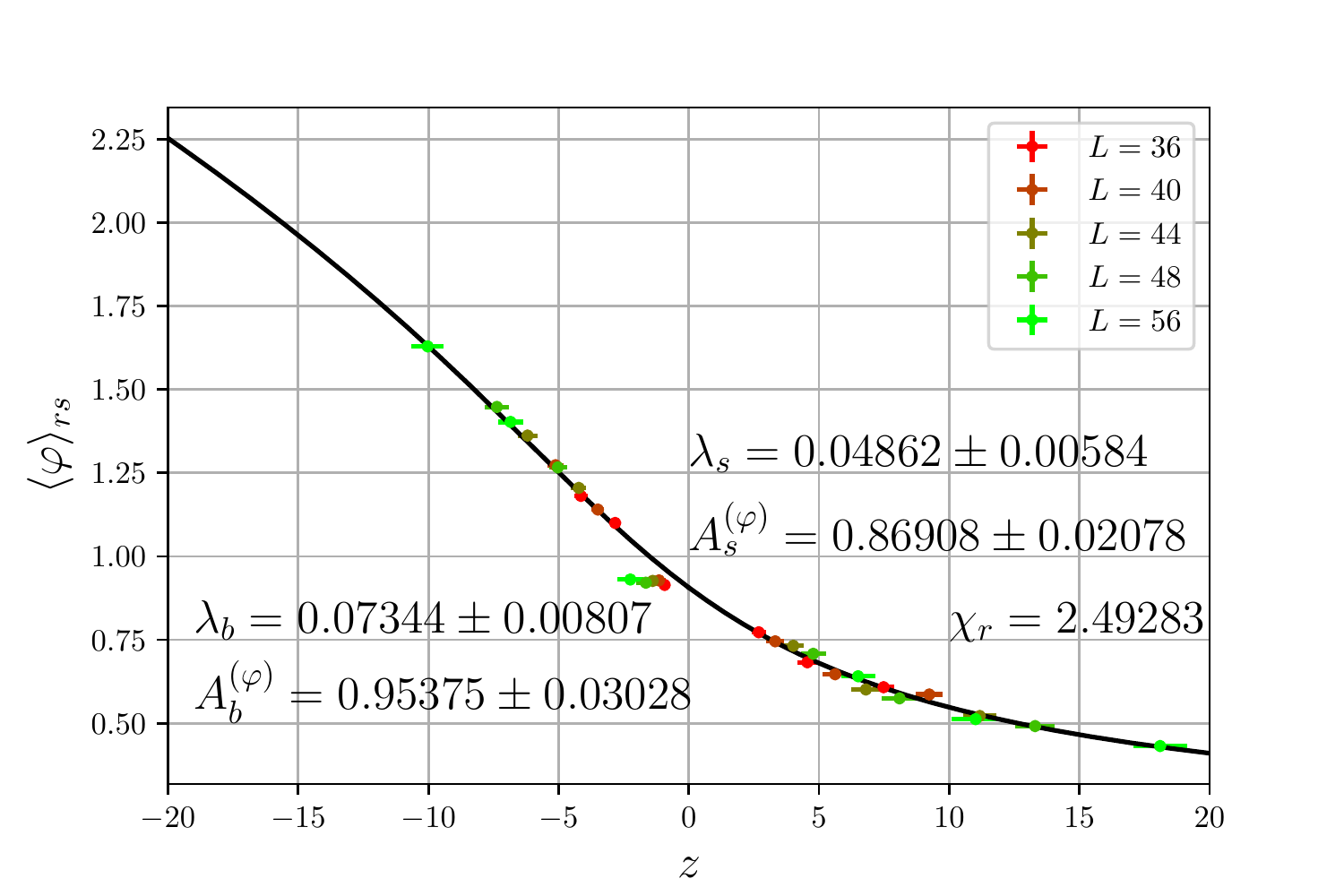}
  \includegraphics[width=4.65cm,clip]{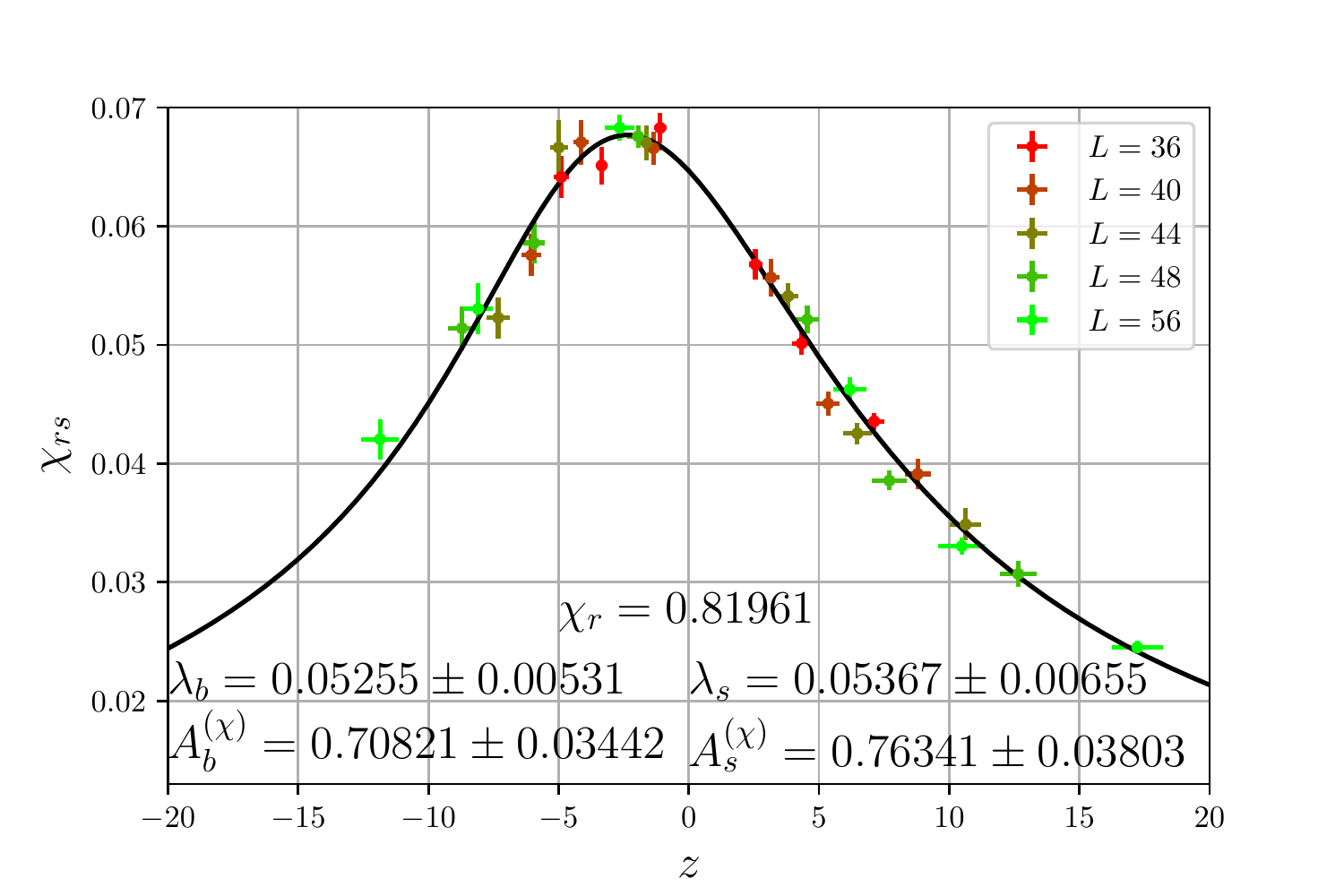}
  \includegraphics[width=4.65cm,clip]{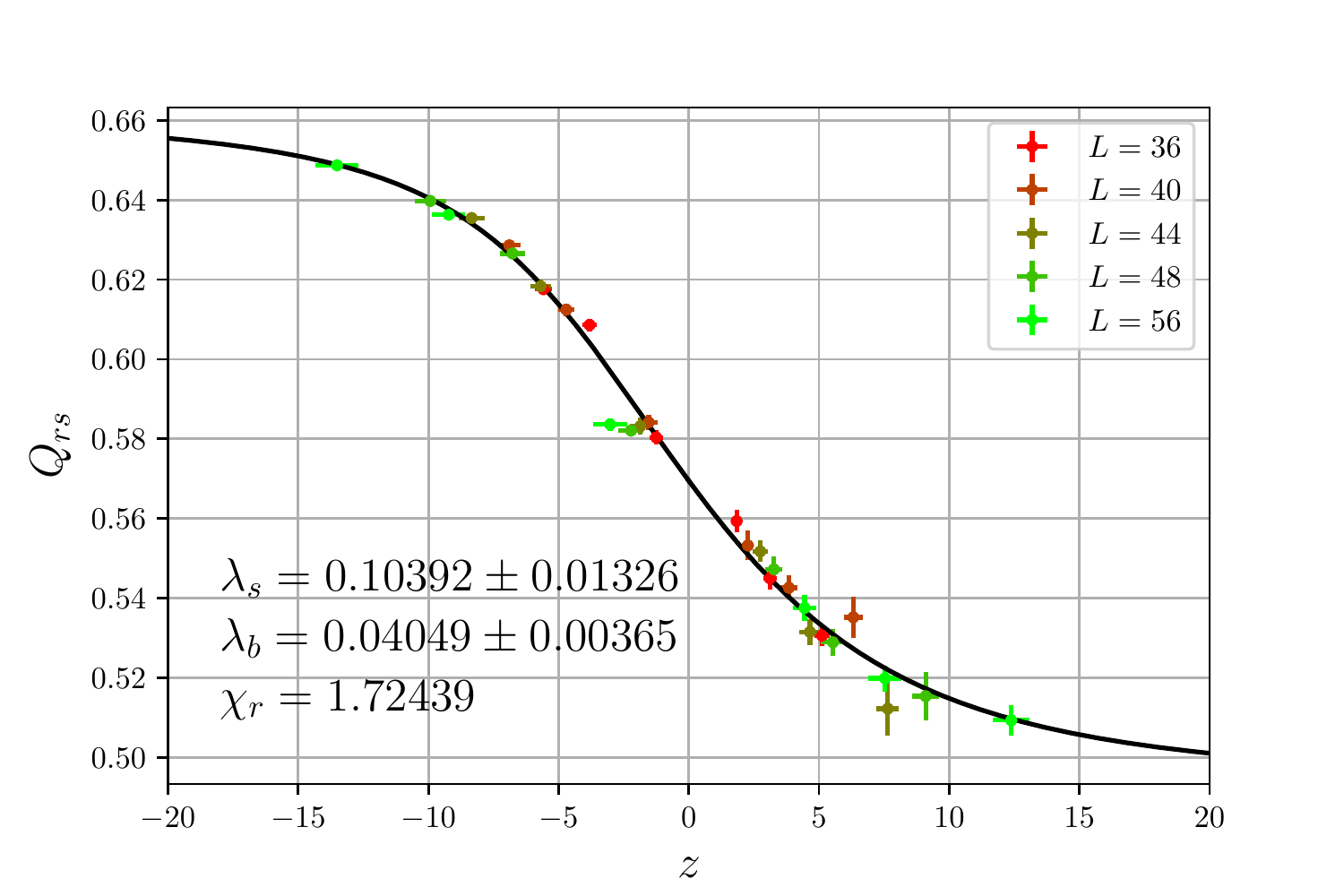}
  \caption{Scaling behaviour of the Higgs-field vev, its
    susceptibility and Binder's cumulant in the pure-scalar $O(4)$ model.  All dimensionful
  quantities are expressed in lattice units.  The subscript $rs$ means
these quantities are rescaled properly to remove the volume
dependence introduced {\it via} the effect of  wavefunction
renormalisation.  The parameters
  $\lambda_{s,b}$ are $\lambda_{m_{R}}$ in the symmetric and the
  broken phases, with similar symbols indicating the relevant
  $A^{(\varphi)}$ and $A^{(\chi)}$ in these two phases, and $\chi_{r}$
indicates the $\chi^{2}/{\mathrm{d.o.f.}}$ of the fit.}
  \label{fig:Plot_scaling}
\end{figure}

Figure~\ref{fig:Fit_scaling} shows the results of the fits for
$\langle \varphi \rangle$, $\chi$ and $Q$ in the pure-scalar $O(4)$
model using the scaling formulae in 
Eqs.~(\ref{eq:scaling_formulae}) and~(\ref{eq:lambda_R_O4}).  The fit
parameters are $\lambda_{m_{R}}$ and $A^{(\varphi, \chi)}$.  We
distinguish them in the symmetric and the broken phases, since the
numerical values of these parameters need not be the same in two
different phases.  Upon extracting these parameters from our lattice
numerical result, we can then evaluate the scaling variable, $z$, for
our data points, as well as removing (rescaling away) the volume-dependence introduced
{\it via} the effect of the wavefunction renormalisation in
Eq.~(\ref{eq:scaling_formulae}).  This allows us to test our
finite-size scaling formulae through plotting the rescaled $\langle
\varphi \rangle$, $\chi$ and $Q$ as functions of $z$.  In
Fig.~\ref{fig:Plot_scaling}, it is demonstrated that these rescaled
quantities all lie on universal curves that are only functions of
$z$.  Such behaviour, together with the good or reasonable values of
the $\chi^{2}$, show strong evidence that our formulae indeed
capture the scaling properties of the theory as governed by the
Gaussian fixed point.  

\section{Conclusion and outlook}
In this article, we present two projects on lattice simulations for the
Higgs-Yukawa model.   The results from our study show that
Lattice Field Theory can be employed for investigating
non-perturbative aspects of the model.

Our first project is the search for phenomenologically viable scenarios
for a strong first-order thermal phase transition in the Higgs-Yukawa
theory with the addition of a dimension-six operator.  This
dimension-six operator can serve as a prototype of new physics.  In
this work, we demonstrate that such first-order transitions can indeed
be observed, when the cut-off scale is kept high in comparison to the
Higgs-boson vev.  However, we are yet to find a suitable choice of
parameters which leads to large enough Higgs-boson mass.  Currently we
are performing more lattice computations to further scan the bare
parameter space.

The second project presented in this article is our work on
finite-size scaling behaviour of the Higgs-Yukawa model near the
Gaussian fixed point.   In this regard, we have derived the scaling
formulae by solving the path integrals at one-loop.  These formulae
are tested against lattice numerical data in the pure-scalar $O(4)$
model, where good agreement is found.  Such formulae can be important
tools for future works on confirming the triviality of the Higgs-Yukawa theory, or 
searching for alternative scenarios where strong-coupling fixed points exist.

\section*{Acknowledgments}
DYJC and CJDL acknowledge the Taiwanese MoST grant number
105-2628-M-009-003-MY4.  DYJC also thanks the financial support from
Chen Cheng Foundation.  

%\clearpage
\bibliography{lattice2017}

\begin{thebibliography}{25}

\bibitem{Hasenfratz:1992xs}
A.~Hasenfratz, K.~Jansen, Y.~Shen, Nucl. Phys. \textbf{B394}, 527 (1993),
  \texttt{hep-lat/9207006}

\bibitem{Gerhold:2007yb}
P.~Gerhold, K.~Jansen, JHEP \textbf{09}, 041 (2007), \texttt{0705.2539}

\bibitem{Gerhold:2007gx}
P.~Gerhold, K.~Jansen, JHEP \textbf{10}, 001 (2007), \texttt{0707.3849}

\bibitem{Bulava:2012rb}
J.~Bulava, P.~Gerhold, K.~Jansen, J.~Kallarackal, B.~Knippschild, C.J.D. Lin,
  K.I. Nagai, A.~Nagy, K.~Ogawa, Adv. High Energy Phys. \textbf{2013}, 875612
  (2013), \texttt{1210.1798}

\bibitem{Grojean:2004xa}
C.~Grojean, G.~Servant, J.D. Wells, Phys. Rev. \textbf{D71}, 036001 (2005),
  \texttt{hep-ph/0407019}

\bibitem{Huang:2015izx}
F.P. Huang, P.H. Gu, P.F. Yin, Z.H. Yu, X.~Zhang, Phys. Rev. \textbf{D93},
  103515 (2016), \texttt{1511.03969}

\bibitem{Damgaard:2015con}
P.H. Damgaard, A.~Haarr, D.~O'Connell, A.~Tranberg, JHEP \textbf{02}, 107
  (2016), \texttt{1512.01963}

\bibitem{Cao:2017oez}
Q.H. Cao, F.P. Huang, K.P. Xie, X.~Zhang (2017), \texttt{1708.04737}

\bibitem{deVries:2017ncy}
J.~de~Vries, M.~Postma, J.~van~de Vis, G.~White (2017), \texttt{1710.04061}

\bibitem{Gockeler:1992zj}
M.~Gockeler, H.A. Kastrup, T.~Neuhaus, F.~Zimmermann, Nucl. Phys.
  \textbf{B404}, 517 (1993), \texttt{hep-lat/9206025}

\bibitem{Bilenky:1994kt}
M.S. Bilenky, A.~Santamaria, \emph{{Beyond the standard model with effective
  lagrangians}}, in \emph{{28th International Symposium on Particle Theory
  Wendisch-Rietz, Germany, August 30-September 3, 1994}} (1994), pp. 215--224,
  \texttt{hep-ph/9503257},
  \urlstyle{tt}\url{https://inspirehep.net/record/382885/files/arXiv:hep-ph_9503257.pdf}

\bibitem{Chu:2015nha}
D.Y.J. Chu, K.~Jansen, B.~Knippschild, C.J.D. Lin, A.~Nagy, Phys. Lett.
  \textbf{B744}, 146 (2015), \texttt{1501.05440}

\bibitem{Fukuda:1974ey}
R.~Fukuda, E.~Kyriakopoulos, Nucl. Phys. \textbf{B85}, 354 (1975)

\bibitem{ORaifeartaigh:1986axd}
L.~O'Raifeartaigh, A.~Wipf, H.~Yoneyama, Nucl. Phys. \textbf{B271}, 653 (1986)

\bibitem{Akerlund:2015fya}
O.~Akerlund, P.~de~Forcrand, Phys. Rev. \textbf{D93}, 035015 (2016),
  \texttt{1508.07959}

\bibitem{Molgaard:2014mqa}
E.~Mølgaard, R.~Shrock, Phys. Rev. \textbf{D89}, 105007 (2014),
  \texttt{1403.3058}

\bibitem{Gies:2009hq}
H.~Gies, M.M. Scherer, Eur. Phys. J. \textbf{C66}, 387 (2010),
  \texttt{0901.2459}

\bibitem{Frohlich:1982tw}
J.~Frohlich, Nucl. Phys. \textbf{B200}, 281 (1982)

\bibitem{Luscher:1988uq}
M.~Luscher, P.~Weisz, Nucl. Phys. \textbf{B318}, 705 (1989)

\bibitem{Hogervorst:2011zw}
M.~Hogervorst, U.~Wolff, Nucl. Phys. \textbf{B855}, 885 (2012),
  \texttt{1109.6186}

\bibitem{Siefert:2014ela}
J.~Siefert, U.~Wolff, Phys. Lett. \textbf{B733}, 11 (2014), \texttt{1403.2570}

\bibitem{Chu:2015jba}
D.Y.J. Chu, K.~Jansen, B.~Knippschild, C.J.D. Lin, K.I. Nagai, A.~Nagy, PoS
  \textbf{LATTICE2015}, 230 (2016), \texttt{1510.08620}

\bibitem{Chu:2016svq}
D.Y.J. Chu, K.~Jansen, B.~Knippschild, C.J.D. Lin, A.~Nagy, PoS
  \textbf{LATTICE2016}, 217 (2016), \texttt{1611.00466}

\bibitem{Brezin:1985xx}
E.~Brezin, J.~Zinn-Justin, Nucl. Phys. \textbf{B257}, 867 (1985)

\bibitem{our_scaling_paper}
D.Y.J. Chu, K.~Jansen, B.~Knippschild, C.J.D. Lin, in preparation

\end{thebibliography}

%%%%%%%%%%%%%%%%%%%%%%%%%%%%%%%%%%%%%%%%%%%%%%%%%%%%%%%%%%%%%%%%%%%%%%%%%%%%%
\end{document}